\documentstyle[twoside,fleqn,espcrc2]{article}

\newcommand{\AmS}{{\protect\the\textfont2
A\kern-.1667em\lower.5ex\hbox{M}\kern-.125emS}}
\title{Non-Perturbative Renormalisation and Kaon Physics}
\author{Massimo Testa\address{Dipartimento di Fisica, Universita' di Roma "La
Sapienza",\\ P.le A.Moro 2, 00185 Roma,Italy\\ INFN-Sezione di Roma, Italy}}
\begin{document}
\begin{abstract} A general review is presented on the problem of non
perturbative computation of the
$K\to\pi\pi$ transition amplitude.
\end{abstract}
\maketitle
\section{INTRODUCTION}
Kaon Physics is a very complicated blend of Ultraviolet and Infrared effects.

The physical problem is the large enhancement shown in $K$-decays:
\begin{eqnarray}
& & {{\Gamma (K^+\to \pi ^+\pi ^0)} \over {\Gamma (K_s^0\to \pi ^+\pi
^-)}}\approx {1 \over {400}}\approx {{\left| A(\Delta I={3 \over 2})
\right|^2} \over {\left| A(\Delta I={1 \over 2}) \right|^2}} \Rightarrow \nonumber\\
& & \Rightarrow A(\Delta I={1 \over 2})\approx 20A(\Delta I={3 \over 2}) \end{eqnarray}

Due to the difficulty of putting the Standard Model on the lattice
for technical ($M_{W,Z} >>{1\over a}$) and theoretical reasons (the
presence of the Wilson term breaks the chiral gauge invariance) we must
ÒintegrateÓ analytically the heavy degrees of freedom ($W^{'}$s, $Z^{'}$s).

Using second order Weak Interaction perturbation theory, we get for the
effective low energy hamiltonian density of non-leptonic decays:
\begin{eqnarray}
& & H_{\Delta S=1}^{eff}=g_W^2\int {d^4xD(x;M_W)T[J_L^\mu (x)J_{L\mu }^+(0)]}+\nonumber\\
& & +c_mO_m \label{one}
\end{eqnarray}
where $D(x;M_W)$ denotes the free $W$ propagator:
\begin{equation}
D(x;M_W)\equiv \int {{{d^4p} \over {(2\pi )^4}}{{\exp (ipx)}
\over {(p^2+M_W^2)}}}\label{two}
\end{equation}
and:
\begin{equation}
O_m\equiv (m_s+m_d)\bar sd+(m_s-m_d)\bar s\gamma _5d+h.c.\label{three}
\end{equation}
$c_m$ is fixed, in Perturbation Theory, so that quark masses are not
modified by Weak Interactions.

The term proportional to $O_m$ does not contribute to the physical
transition amplitude (a matrix element with zero four momentum transfer, 
$\Delta p_\mu =0$), being a total four-divergence.

Since $M_W>>\Lambda _{QCD}$ we can use the Operator Product Expansion, to
get:
\begin{equation}
H_{\Delta S=1}^{eff}=\lambda _u{{G_F} \over {\sqrt 2}}\left[ {C_+(\mu)O^{(+)}(\mu )+
C_-(\mu)O^{(-)}(\mu )} \right]\label{four}
\end{equation}
where $\lambda _u=V_{ud}V_{us}^*$ and:
\begin{equation}
O^{(\pm )}=\left[ {(\bar s\gamma _\mu ^Ld)(\bar u\gamma _\mu ^Lu)\pm (\bar
s\gamma _\mu ^Lu)(\bar u\gamma _\mu ^Ld)} \right]-\left[ {u\leftrightarrow c}
\right]\label{five}
\end{equation}
In Eq.[\ref{four}] the subtraction point
$\mu $ is chosen so that $\mu >m_c$  and we are in the favorable
situation of propagating charm with the consequent GIM
cancellation\cite{ottoa}.
$O^{(-)}$ is an operator contributing to pure $\Delta I={1 \over 2}$
transitions, while
$O^{(+)}$ contributes both to
$\Delta I={1 \over 2}$ and $\Delta I={3 \over 2}$ transitions.
The $C_\pm $, computed in Perturbation Theory, show a slight
enhancement\cite{uno}:
\begin{equation}
\left| {{{C_-(\mu \approx 2\;Gev)} \over {C_+(\mu \approx 2\;Gev)}}}
\right|\approx 2
\end{equation}

The rest of the enhancement ($\approx 10$) should be provided by the matrix
elements of $O^{(\pm )}$ and is a non perturbative, infrared effect.

In this talk I will be concerned with the Wilson-like fermion formulation.
Several interesting papers deal with problems related to non-octet
$K$-transitions, both within the Wilson and staggered fermion
approaches\cite{due}.

The difficulty of the problem consists, first of all, in giving the correct
ultraviolet convergent definition of the operators $O^{(\pm )}$.
It is well known that, in order to construct a finite composite operator of
dimension 6, $\tilde O_6(\mu )$, we must mix the original bare operator,
$O_6(a)$, with bare operators of equal ($O_6^{(i)}(a)$) or smaller
($O_3(a)$) dimension, in general with different naive chiralities\cite{unoa}.

In the present case we have, schematically:
\begin{eqnarray}
& &\tilde O_6(\mu )=Z(\mu a)\times \label{six} \\
&& \times [{O_6(a)+\sum\limits_i
C_i(g_0)O_6^{(i)}(a)+{{C_3(g_0)}\over{a^3}}O_3(a)}]\nonumber
\end{eqnarray}

Being QCD asymptotically free, one could think that the mixing coefficients in Eq.[\ref{six}]
could be reliably computed in Perturbation Theory.

Numerical attempts (and theoretical considerations) show that this is not
necessarily the case. We are therefore led to employ general non perturbative techniques.
On very general grounds, these techniques are based on the systematic exploitation of (continuum) symmetries.
In fact it turns out that the $O^{(\pm )}{'}$s have definite transformation properties under the $SU(3)\otimes SU(3)$ chiral group (and some discrete
symmetries).

This fact suggests the use of Chiral Ward Identities (or equivalent methods) to
classify the composite operators\cite{unoa},\cite{tre}.

The Chiral Ward Identities, however, do not fix in a completely unambiguous
way the $O^{(\pm ){'}}$s.
Some ambiguities are left, outside of the chiral limit. These ambiguities turn out to be irrelevant
in cases in which Current Algebra may be applied. They, however, limit the application of the
method to light meson decays. In order to treat other interesting physical processes, involving e.g. B-mesons, other methods must be used, as discussed later.

\section{ULTRAVIOLET PROBLEMS}
\subsection{The Chiral Limit}
As already stated, in order to construct an operator with the correct
(continuum) chirality, we must mix it with operators $O_\alpha ^{(i)}$ of
equal or smaller dimension, with different naive chiralities:
\begin{equation}
\tilde O_\alpha =O_\alpha +\sum\limits_i {c_iO_\alpha ^{(i)}}\label{seven}
\end{equation}

A general way to see why Eq.[\ref{seven}] is correct, is to think of the lattice
discretization of QCD as a chiral violating order $a$ perturbation of the
continuum field theory, described by an action:
\begin{eqnarray}
& & S_{QCD}=S_{cont}+\nonumber\\
& & +\int {d^4y\bar q(y)mq(y)+}a\int {d^4yW(y)}\label{eight}
\end{eqnarray}
where $S_{cont}$ describes the continuum theory, in the chiral limit, and
$W(y)$ is a chiral violating, dimension 5 operator whose origin can be traced
back to the presence, in the Lattice Action, of the Wilson term.
The existence of $W(y)$ forces us, already in the chiral limit, $m=0$, to
perform a complicated subtraction procedure in order to define finite ultraviolet composite operators with good chiral transformation properties.

Formally we may write:
\begin{eqnarray}
& & \left\langle A \right|O_\alpha (0)\left| B \right\rangle
_{(lat)}\approx \left\langle A \right|\tilde O_\alpha (0)\left| B
\right\rangle _{(cont)}+\nonumber\\
& & + \sum\limits_{n\ge 1} {{{a^n} \over {n!}}\left\langle A
\right|T{\tilde O_\alpha (0)  (\int {dyW(y)})^n}\left| B
\right\rangle}_{(cont)}\label{nine}
\end{eqnarray}
where $\tilde O_\alpha (0)$ is, by definition, a finite operator which
satisfies the chiral continuum (integrated) Ward Identities:
\begin{eqnarray}
& & \mathop {\lim }\limits_{m\to 0}\int {dz\left\langle A
\right|T({\delta _{op}[\bar q(z)mq(z)]\tilde O_\alpha (0)})\left| B
\right\rangle _{(cont)}}+\nonumber\\
& & +\left\langle A \right|\delta _{op}\tilde O_\alpha (0)\left| B
\right\rangle _{(cont)}=0\label{ten}
\end{eqnarray}
where $\delta _{op}$ denotes the operator chiral variation.
The Ward Identity Eq.[\ref{ten}] is written in a very general  form and can be applied both to the case in which chiral symmetry is restored in the massless limit, and to the case, realized in QCD, where chiral symmetry is spontaneously broken as $m\to 0$. In this case the mass insertion term in Eq.[\ref{ten}] does not vanish in the chiral limit because of the presence of the pion pole. 
It is to be noted that no prescription is needed for the (generally divergent) insertion of
$\delta _{op}[\bar q(z)mq(z)]$ and $\tilde O_\alpha (0)$, since we are considering the limit $m\to 0$ and the divergent part does not
contain the pion pole.
	
The $a$ dependence cannot be directly read from Eq.[\ref{nine}]
because of the presence of multiple insertions of composite operators
$\tilde O_\alpha (0)$ and $W(y)$ which are, in general, divergent.

For example for $n=1$ we have:
\begin{eqnarray}
& & \left\langle A \right|O_\alpha (0)\left| B \right\rangle
_{(lat)}\approx \left\langle A \right|\tilde O_\alpha (0)\left| B
\right\rangle _{(cont)}+\nonumber\\
& & +a\int {dy\left\langle A \right|T({\tilde O_\alpha (0)W(y)})\left| B
\right\rangle _{(cont)}}\label{eleven}
\end{eqnarray}
The (continuum) Operator Product Expansion implies (up to log corrections):
\begin{equation}
T(\tilde O_\alpha (0)W(y))\mathop \approx \limits_{y\approx 0}\sum\limits_i
{{1 \over {y^{(5+d_O-d_{O^{(i)}})}}}O_\alpha ^{(i)}(0)}\label{twelve}
\end{equation}
Because of the multiplying $a$ factor, in Eq.[\ref{eleven}] only the short distance behavior contributes at leading order in $a$:
\begin{eqnarray}
& & \left\langle A \right|O_\alpha (0)
\left| B\right\rangle_{(lat)}\approx\nonumber\\
& & \approx \left\langle A \right|\tilde O_\alpha (0)\left| B 
\right\rangle_{(cont)}+\label{thirteen}\\
& & +\sum\limits_i {{1\over{a^{(d_O-d_{O^{(i)}})}}}\left\langle A
\right|O_\alpha ^{(i)}(0)\left| B \right\rangle _{(cont)}}\nonumber
\end{eqnarray}

The $O_\alpha ^{(i)}$Ôs in Eqs.[\ref{twelve}],[\ref{thirteen}] are operators which generally belong to chiral
representations different from that of $O_\alpha $ (apart from trivial multiplicative renormalisation).
In the Chiral Limit the Ward Identities completely determine the structure
of the composite operator $\tilde O_\alpha $.

In fact Eq.[\ref{thirteen}] implies the mixing structure which has the form  indicated in Eq.[\ref{seven}]:
\begin{eqnarray}
& & \left\langle A \right|\tilde O_\alpha (0)\left| B \right\rangle
_{(cont)}\approx \left\langle A \right|O_\alpha (0)\left| B \right\rangle
_{(lat)}+\\
& & -\sum\limits_i {{1 \over {a^{(d_O-d_{O^{(i)}})}}}\left\langle A
\right|O_\alpha ^{(i)}(0)\left| B \right\rangle _{(lat)}}+...\nonumber
\end{eqnarray}

$\tilde O_\alpha $ satisfies the Ward Identities up to $O(a)$.

We also see that the leading divergent part of the mixing,
${1 \over{a^{(d_O-d_{O^{(i)}})}}}$, is determined by perturbation theory
(short distance).
\subsection{Softly Broken Chiral Symmetry}

When the explicit chiral symmetry breaking is switched on
($m\ne 0$), other divergences (and ambiguities) may arise.

This happens also in the continuum.

In order to discuss the case $m\ne 0$ it is more suitable to think of a
generalized mass term of the form $\bar q_{Ri}M_{ij}q_{Lj}+h.c.$
(in the end we shall, of course, choose $M$ as an hermitian diagonal matrix).

If we transform:
\begin{eqnarray}
& & q_{Li}\to L_{ik}q_{Lk}\nonumber\\
& & q_{Ri}\to R_{ik}q_{Rk}\label{fourteen}
\end{eqnarray}
and, simultaneously:
\begin{equation}
M\to RML^+\label{fifteen}
\end{equation}
then the mass insertion remains invariant.

Up to first order in $M$, we have:
\begin{eqnarray}
& & \left\langle A \right|\tilde O_\alpha (0)\left| B \right\rangle
_{(M)}\approx\left\langle A \right|\tilde O_\alpha (0)\left| B \right\rangle
_{(M=0)}+\nonumber\\
& & +\int {dy\left\langle A \right|T{\tilde O_\alpha (0)\bar
q(y)Mq(y)}\left| B \right\rangle _{(M=0)}}\label{sixteen}
\end{eqnarray}

As usual the double insertion of composite operators gives rise to a
divergent contribution determined by the operator product expansion:
\begin{eqnarray}
& & T({\tilde O_\alpha (0)\;\bar q(y)Mq(y)})\mathop \approx \limits_{y\approx
0}\nonumber\\
& & \approx \sum\limits_i {{1 \over {y^{(3+d_{\tilde O}-d_{A_i})}}}(MA)_\alpha
^{(i)}(0)}\label{eighteen}
\end{eqnarray}
where the operators $A_\alpha ^i(0)$ have chiralities different from that of
$\tilde O_\alpha (0)$. $(MA)_\alpha ^{(i)}$ denotes a suitable combination of
the mass parameters $M$ and the operators $A_\alpha ^i(0)$.

Under a simultaneous chiral rotation, Eq.[\ref{fourteen}], on both sides of
Eq.[\ref{eighteen}] and the corresponding transformation Eq.[\ref{fifteen}], we see that the
operator combination
$(MA)_\alpha ^{(i)}$ transforms according the same chiral representation as
$\tilde O_\alpha (0)$, since $\bar qMq$ is invariant\cite{quattro}:
\begin{equation}
\delta (MA)_\alpha ^{(i)}=(\delta M\;A)_\alpha ^{(i)}+(M\;\delta
_{op}A)_\alpha ^{(i)}\approx \delta _{op}\tilde O_\alpha\label{nineteen}
\end{equation}

We can now define the finite operator:
\begin{equation}
\tilde {\tilde O}_\alpha (0)\equiv \tilde O_\alpha (0)-\sum\limits_i {{1
\over {a^{(d_{\tilde O}-d_{A_i}-1)}}}(MA)_\alpha ^{(i)}(0)}\label{twenty}
\end{equation}
and study its Ward Identities.

We start with the (divergent, but regularised) Ward Identity:
\begin{eqnarray}
& & \int {dy\left\langle A \right|T({M_{ij}\delta _{op}[\bar
q_i(y)q_j(y)]\tilde {\tilde O}_\alpha (0)})\left| B \right\rangle
}+\nonumber\\
& & +\left\langle A \right|\delta _{op}\tilde {\tilde O}_\alpha (0)\left| B
\right\rangle =\nonumber\\
& & =-\int {dy\left\langle A \right|T({\bar q(y)(\delta \kern
1pt M)q(y)\tilde{\tilde O}_\alpha (0)})\left| B \right\rangle}+\nonumber\\
& & +\left\langle A
\right|\delta _{op}\tilde {\tilde O}_\alpha (0)\left| B \right\rangle =0\label{twentyone}
\end{eqnarray}

Divergent contributions are present in Eq.[\ref{twentyone}]. They arise from:

1) the double operator insertion in the $T$-product

2) since $\delta _{op}$ denotes the operator chiral variation, $\tilde
{\tilde O}_\alpha (0)$ transforms as a superposition of the representations
of $\tilde O_\alpha (0)$ and of $A_\alpha ^i(0)$:
\begin{eqnarray}
& & \delta _{op}\tilde {\tilde O}_\alpha (0)\equiv \delta _{op}
{\tilde O}_\alpha(0)+\label{twentytwo}\\
& & -\sum\limits_i {{1 \over
{a^{(d_{\tilde {O}}-d_{A_i}-1)}}}(M\delta _{op}A)_\alpha ^{(i)}(0)}\nonumber
\end{eqnarray}
so that the single insertion of $\delta _{op}\tilde {\tilde O}_\alpha (0)$ is
also divergent.

In order to make everything finite we may redefine a $T^*$-product
according to the prescription:
\begin{eqnarray}
& & T^*({\tilde {\tilde O}_\alpha (0)\;\bar q(y)Mq(y)})=\label{twentythree}\\
& & =T({\tilde {\tilde O}_\alpha (0)\;\bar q(y)Mq(y)})+\nonumber\\
& & -\sum\limits_i {{{\delta
^4(y)} \over {a^{(d_{\tilde O}-d_{A_i}-1)}}}(MA)_\alpha ^{(i)}(0)}\nonumber
\end{eqnarray}

In this way we rewrite the Ward Identity Eq.[\ref{twentyone}] as:
\begin{eqnarray}
& & -\int {dy\left\langle A \right|T^*{\bar q(y)(\delta \kern 1pt
M)q(y)\tilde {\tilde O}_\alpha (0)}\left| B \right\rangle }+\nonumber\\
& & +\left\langle A \right|\delta _{op}\tilde {\tilde O}_\alpha (0)\left| B
\right\rangle+\label{twentyfour}\\
& &  -\sum\limits_i {{1 \over {a^{(d_{\tilde
O}-d_{A_i}-1)}}}\left\langle A \right|(\delta \kern 1pt MA)_\alpha
^{(i)}(0)}\left| B \right\rangle=0\nonumber
\end{eqnarray}

From:
\begin{equation}
-(\delta M\;A)_\alpha ^{(i)}=(M\;\delta _{op}A)_\alpha ^{(i)}-\delta
(MA)_\alpha ^{(i)}\label{twentyfive}
\end{equation}
we then get:
\begin{eqnarray}
& & -\int {dy\left\langle A \right|T^*{\bar q(y)(\delta 
M)q(y)\tilde {\tilde O}_\alpha (0)}\left| B \right\rangle }+\nonumber\\
& & +\left\langle A
\right|\delta _{op}\tilde {\tilde O}_\alpha (0)\left| B
\right\rangle+\label{twentysex}\\
& & +\sum\limits_i {{1 \over {a^{(d_{\tilde
O}-d_{A_i}-1)}}}\left\langle A \right|(M\delta _{op}A)_\alpha
^{(i)}(0)}\left| B \right\rangle+\nonumber\\
& & -\sum\limits_i {{1 \over {a^{(d_{\tilde O}-d_{A_i}-1)}}}\left\langle A
\right|\delta (MA)_\alpha ^{(i)}(0)\left| B \right\rangle }=0\nonumber
\end{eqnarray}
Eq.[\ref{twentysex}] is easily seen to be equivalent to:
\begin{eqnarray}
& & \int {dy\left\langle A \right|T^*({M_{ij}\delta _{op}(\bar
q_i(y)q_j(y))\;\tilde {\tilde O}_\alpha (0)})\left| B\right\rangle}
+\nonumber\\
& & +\left\langle A \right|\delta\;\tilde {\tilde O}_\alpha
(0)\left| B\right\rangle =\nonumber\\
& & =-\int {dy\left\langle A \right|T^*({\bar
q(y)(\delta \kern 1pt M)q(y)\;\tilde {\tilde O}_\alpha (0)})\left| B
\right\rangle }+\nonumber\\
& & +\left\langle A
\right|\delta \;\tilde {\tilde O}_\alpha (0)\left| B \right\rangle =0\label{twentyseven}
\end{eqnarray}
So that, in the end, with a joint redefinition of the operator and the $T$-product
containing the mass insertion, we managed to have a renormalised Ward
Identity of the canonical form.
It is now clear why the Ward Identity does not uniquely define the finite
operator $\tilde {\tilde O}_\alpha (0)$, outside the chiral limit.
In fact we may change the definition of the $T^*$-product and accordingly
change the definition of the operator $\tilde {\tilde O}_\alpha (0)$ by a
finite mixing with $(MA)_\alpha ^{(i)}$, leaving the form of the Ward Identity
unaltered.

This ambiguity is harmless if we are in a position to apply the
low-energy theorems of Current Algebra (small quark masses) or when the
physical matrix elements of $(MA)_\alpha ^{(i)}$ are zero (e.g. if they are
proportional to four divergences of current), but it can represent a serious obstruction when
heavy hadrons are present.

We conclude this section by remarking that the situation presented here is a bit simplified.
In fact we should also consider multiple insertions of $\bar q(x)Mq(x)$
and $W(y)$, which, in general, generate further distortions. Such terms will be included in the
following considerations.

\subsection{Non Perturbative Renormalisation}

A different, but equivalent construction of composite operators (in
agreement with the Ward Identities for physical applications) can be achieved
through the so called Non Perturbative Renormalisation\cite{cinque}.

This method proceeds in complete parallelism with the corresponding
procedure used in perturbation theory.

In fact non perturbative renormalisation requires a non-perturbative gauge
fixing procedure, usually in the form of a suitable discretization of the Landau gauge
$\partial ^\mu G_\mu =0$ \footnote{Gribov copies should not present a
serious problem in the present case, because gauge fixing is only used in the
computation of short distance quantities which are probably immune from their
presence.}.

The basic idea is to consider the insertion of the renormalised operator
$\tilde {\tilde O}_\alpha =O_\alpha +\sum\limits_i {c_iO_\alpha ^{(i)}}$
in Green's functions containing elementary quark (and gluon) fields.

Symbolically we write:
\begin{equation}
G_{\tilde {\tilde O}}(p)\equiv F.T.\;\left\langle {\tilde {\tilde O}_\alpha
(0)q(x)\bar q(y)} \right\rangle \label{twentyeight}
\end{equation}
where $F.T.$ denotes the appropriate Fourier Transform of the external legs.

We expect that, for large (continuum) virtualities ${1 \over a}>>p\approx \mu
>>\Lambda _{QCD}$, the chiral violating form factors of $G_{\tilde {\tilde
O}}(p)$ (which come from physical soft mass breaking and/or spontaneous
symmetry breaking) should be suppressed by inverse powers of $\mu$.

As an example we could consider the correlator
$F.T.\;\left\langle {s(x)\;\bar s_Rd_L(0)\;\bar d(y)} \right\rangle$.
In this case the conditions are:
\begin{eqnarray}
& & F.T.\;\left\langle {s_L(x)\;\bar s_Rd_L(0)\;\bar d_{L,R}(y)}
\right\rangle _{\mu \to \infty }=0\nonumber\\
& & F.T.\;\left\langle {s_{L,R}(x)\;\bar s_Rd_L(0)\;\bar d_R(y)} \right\rangle
_{\mu \to \infty }=0\label{twentynine}
\end{eqnarray}

This behavior is a direct consequence of the (integrated) Chiral Ward
Identity. I will discuss, for simplicity, the case in which mixing
with operators of lower dimension do not occur. This case is directly relevant
for the study of the transitions $K^+\to \pi ^+\pi ^0$ and $K^0\leftrightarrow
\bar K^0$.

The integrated Ward Identity has the form:
\begin{eqnarray}
& & F.T.\;\int {dz\left\langle {\delta _{op}[\bar q(z)Mq(z)]\;\tilde
{\tilde O}_\alpha (0)q(x)\bar q(y)} \right\rangle }+\nonumber\\
& & +F.T.\;\left\langle {\delta \;\tilde {\tilde O}_\alpha (0)q(x)\bar q(y)}
\right\rangle +\nonumber\\
& & +F.T.\;\left\langle {\tilde {\tilde O}_\alpha (0)\;\delta \,q(x)\bar q(y)}
\right\rangle +\nonumber\\
& & +F.T.\;\left\langle {\tilde {\tilde O}_\alpha (0)q(x)\;\delta \,\bar q(y)}
\right\rangle =0\label{thirty}
\end{eqnarray}

Because of the explicit $M$ factor, the l.h.s. of Eq.[\ref{thirty}] has
one {\em operator} dimension less than the individual terms appearing on
the r.h.s. so that, at large virtualities, it vanishes one power
faster. 
For asymptotically large $\mu$ we then get from Eq.[\ref{thirty}]:
\begin{eqnarray}
& & F.T.\;\left\langle {\delta \;\tilde {\tilde O}_\alpha (0)q(x)\bar
q(y)} \right\rangle _\mu +\nonumber\\
& & +F.T.\;\left\langle {\tilde {\tilde O}_\alpha (0)\;\delta \,q(x)\bar q(y)}
\right\rangle _\mu +\nonumber\\
& & +F.T.\;\left\langle {\tilde {\tilde O}_\alpha (0)q(x)\;\delta \,\bar q(y)}
\right\rangle _\mu \equiv \nonumber\\
& & \equiv F.T.\;\left\langle {\delta \;[\tilde {\tilde O}_\alpha (0)q(x)\bar
q(y)]} \right\rangle _\mu =0 \label{thirtyone}
\end{eqnarray}
which is equivalent to the Wigner-Eckart Theorem (at large virtualities):
\begin{equation}
F.T.\,\left\langle 0 \right|\,\left[ {Q^a,\tilde {\tilde O}_\alpha
(0)q(x)\bar q(y)} \right]\,\left| 0 \right\rangle =0\label{thirtytwo}
\end{equation}

Thus, adjusting the mixing coefficients $c_i$ so that the chiral violating
form factors vanish asymptotically, we get an answer equivalent to the
application of the (softly broken) chiral Ward Identities.

Apart from its simplicity, a further advantage of the non perturbative
renormalisation approach is that it can also determine the absolute
normalization of $\tilde {\tilde O}_\alpha (0)$, corresponding to the
one adopted in perturbation theory, by imposing the same kind of conditions
as, for example:
\begin{equation}
F.T.\;\left. {\left\langle {q\,q\,\bar q\,\bar q\,\tilde {\tilde O}_\alpha }
\right\rangle } \right|_\mu =1 \label{thirtythree}
\end{equation}

\section{INFRARED PROBLEMS}

In order to compute the $K\to \pi \pi $ width we have to evaluate the matrix
element ${}_{(out)}\left\langle {\pi (\underline p)\pi (-\underline p)}
\right|H_W\left| K \right\rangle $ with two interacting hadrons in the final
state.

This is not easy to do in the euclidean region

In fact it can be shown\cite{sei} that the usual strategy of taking large
time limits of euclidean correlators does not give direct information on ${}_{(out)}\left\langle {\pi (\underline p)\pi (-\underline p)}
\right|H_W\left| K \right\rangle $, but is contaminated by final state
interaction effects:
\begin{eqnarray}
& & \left\langle {\pi _{\underline p}(t_1)\pi _{-\underline
p}(t_2)H_W(0)K(t_K)} \right\rangle\approx\label{thirtyfour}\\
& & \mathop \approx \limits_{\scriptstyle {t_K\to
-\infty}\hfill\atop\scriptstyle {t_1>>t_2>>0}\hfill}{{e^{m_Kt_K-E_\pi
t_1-E_\pi t_2}} \over 2}\nonumber\\
& & ({}_{(out)}\left\langle {\pi (\underline
p)\pi (-\underline p)}\right|H_W\left| K \right\rangle +\nonumber\\
& & +{}_{(in)}\left\langle {\pi (\underline p)\pi (-\underline p)}
\right|H_W\left| K \right\rangle +\nonumber\\
& & +e^{2(E_\pi -m_\pi )t_2}cM_{\mathop {\pi \pi }\limits_{2m_\pi }\to \mathop
{\pi \pi }\limits_{2E_\pi }}^{(o.s.)})\nonumber
\end{eqnarray}
where $M_{\mathop {\pi \pi }\limits_{2m_\pi }\to \mathop {\pi \pi
}\limits_{2E_\pi }}^{(o.s.)}$ is the
($\mathop {\pi \pi }\limits_{2m_\pi }\to \mathop {\pi \pi }
\limits_{2E_\pi}$) off-shell scattering amplitude
representing the final state interaction.

If $\underline p=0$ we have:
\begin{eqnarray}
& & \left\langle {\pi _{\underline 0}(t_1)\pi _{\underline
0}(t_2)H_W(0)K(t_K)} \right\rangle \mathop \approx \limits_{\scriptstyle
{t_K\to -\infty }\hfill\atop
  \scriptstyle {t_1>>t_2>>0}\hfill}\label{thirtyfive}\\
& & \approx e^{m_Kt_K-m_\pi t_1-m_\pi t_2}\left\langle {\pi (\underline 0)\pi
(\underline 0)} \right|H_W\left| K \right\rangle \times \nonumber\\
& & \times (1+{c \over {\sqrt
{t_2}}}M_{\mathop {\pi \pi }\limits_{2m_\pi }\to \mathop {\pi \pi}\limits_{2m_\pi}})\nonumber
\end{eqnarray}

Eq.[{\ref{thirtyfive}] shows that only for $\pi^{'}$s at rest (and weakly
interacting) it is possible to extract a meaningful matrix element.

For $K\to \pi \pi $ this could work because chiral symmetry implies a small
final state interaction (Adler zeroÕs). It is however a subject which certainly
deserves further study.

We conclude remarking that the euclidean Green's function for the $\Delta I={1 \over 2}$
transition, defined in Eq.[\ref{thirtyfive}], contains, in general, a disconnected
contribution
\begin{eqnarray}
& & \left\langle {\pi _{\underline 0}(t_1)\pi _{\underline
0}(t_2)H_W(0)K(t_K)} \right\rangle_{disc}=\nonumber \\
& & \left\langle {\pi _{\underline 0}(t_1)\pi _{\underline 0}(t_2)} \right\rangle
\left\langle {H_W(0)K(t_K)}\right\rangle \label{thirtyfivea}
\end{eqnarray}
which must, of course, be subtracted in order to get the
physical result.

\section{HOW TO COMPUTE $K\to \pi \pi $}

This section is based on the considerations exposed in Ref.\cite{sette}.

We will discuss several possible ways to compute $K\to \pi \pi $:

$\bullet$ The Direct Method (and variants thereof)
	(Parity Violating)

$\bullet$ The Parity Conserving Method

$\bullet$ First Principles

\subsection{Direct Methods}

These methods use the strategy first envisaged in Ref.\cite{otto}.
The relevant matrix element is evaluated for $\pi^{'}$s at
rest, which minimizes the final state interactions.

Flavor and $CPS$-symmetry ($CP\otimes (s\leftrightarrow d)$) severely
constrain the form of the subtractions needed to define the Parity
Violating renormalised Weak Hamiltonian density:
\begin{eqnarray}
& & \tilde{\tilde O}_{(P.V.)}^{(\pm)}(\mu)=Z_{(P.V.)}^{(\pm)}(\mu a)\times\label{thirtysix} \\ 
& & \times (O_{(P.V.)}^{(\pm)}(a)+(m_c-m_u){{C_p^{(\pm)}}\over a}O_p(a)) \nonumber
\end{eqnarray}
where:
\begin{equation}
O_p=(m_s-m_d)\bar s\gamma _5d\label{thirtyseven}
\end{equation}

If we choose a world in which $m_s=m_d$ we see from Eq.[\ref{thirtysix}] that
the power divergent mixing with $O_p$ is eliminated.

In this situation no subtractions are needed. The absolute normalization can
be found both perturbatively or non perturbatively, e.g. using external quark
states, as in Eq.[\ref{thirtythree}]\footnote{The disconnected contribution,
Eq.[{\ref{thirtyfivea}}], vanishes by $CPS$ symmetry.}.

We can therefore compute the zero three-momentum transfer matrix element:
\begin{equation}
A_{m_s=m_d}^{(\pm )}=\left. {\left\langle {\pi (\underline 0)\pi
(\underline 0)} \right|O^{(\pm )}(\mu )\left| {K(\underline 0)} \right\rangle
} \right|_{m_s=m_d}\label{thirtyeight}
\end{equation}
and reach the real world through chiral perturbation theory:
\begin{equation}
A_{phys}^{(\pm )}\approx {{m_{K,\;phys}^2-m_{\pi ,\;phys}^2} \over
{2m_{K,\;lat}^2}}A_{m_s=m_d}^{(\pm )}\label{thirtynine}
\end{equation}

Another possible way to eliminate the $O_p$ subtraction in
Eq.[\ref{thirtysix}] is to use a zero four-momentum transfer matrix element.
The basic idea is to work with non perturbatively $O(a)$ improved fermions (i.e.
an S-W improved action with a coefficient $c_{SW}$ determined as in
Ref.\cite{nove}), and choose the quark masses so that $m_K=2m_\pi $.

In the continuum, since we are working at zero four-momentum transfer, we
would have:
\begin{eqnarray}
& & \left\langle h \right|O_p\left| {h'} \right\rangle \propto (m_s+m_d)
\left\langle h\right|\bar s\gamma _5d\left| {h'} \right\rangle =\nonumber\\
 & & =\left\langle h \right|\partial _\mu (\bar s\gamma _\mu \gamma _5d)\left| {h'}\right\rangle
=0\label{forty}
\end{eqnarray}

The computation performed on the non-improved lattice would give:
\begin{eqnarray}
& & {{\left\langle h \right|O_p\left| {h'} \right\rangle } \over
a}\propto (m_s+m_d){{\left\langle h \right|\bar s\gamma _5d\left| {h'}
\right\rangle } \over a}=\label{fortyone}\\
& & ={{\left\langle h \right|\partial _\mu (\bar s\gamma _\mu \gamma
_5d)\left| {h'} \right\rangle } \over a}-{{\left\langle h \right|\bar
X_A\left| {h'} \right\rangle } \over a}=O(1)\nonumber
\end{eqnarray}
and therefore the unsubtracted result would be affected by an error $O(1)$.

Working with improved axial current and pseudoscalar density, we would have:
\begin{eqnarray}
& & {{\left\langle h \right|O_p\left| {h'} \right\rangle } \over
a}\propto (m_s+m_d){{\left\langle h \right|\bar s\gamma _5d\left| {h'}
\right\rangle } \over a}=\nonumber \\
& & ={{\left\langle h \right|\partial _\mu (\bar s\gamma _\mu \gamma
_5d)\left| {h'} \right\rangle } \over a}+{{O(a^2)} \over a}=O(a)\label{fortytwo}
\end{eqnarray}
In this case we may then compute directly:
\begin{equation}
A_{m_K=2m_\pi }^{(\pm )}=\left. {\left\langle {\pi (\underline 0)\pi
(\underline 0)} \right|O_{(P.V.)}^{(\pm )}(\mu )\left| {K(\underline 0)}
\right\rangle } \right|_{m_K=2m_\pi }\label{fortythree}
\end{equation}
without worrying about the subtraction.

The connection with the real world is again obtained through chiral
perturbation theory:
\begin{equation}
A_{phys}^{(\pm )}\approx {{m_{K,\;phys}^2-m_{\pi ,\;phys}^2} \over
{m_{K,\;lat}^2-m_{\pi ,\;lat}^2}}A_{m_K=2m_\pi }^{(\pm )}\label{fortyfour}
\end{equation}

In order to eliminate the disconnected contribution, Eq.[{\ref{thirtyfivea}], we may work with
finite, intermediate renormalised operators obtained by adjusting
$C_p^{(\pm )}$ so that:
\begin{equation}
\left\langle 0 \right|\tilde {\tilde O}_{(P.V.)}^{(\pm )}(\mu )\left| K
\right\rangle =0\label{fortyfive}
\end{equation}
This subtraction does not change the decay rate because it amounts to change the
operator by a four-divergence and reduces numerical errors.

\subsection{Real World}

Still another possibility, based on the proposal of Ref.\cite{dieci} to deal
with final state interactions, is:

1) parametrize the real world $K\to \pi \pi $ amplitude in the Minkowski
region, in a way apt to analytic euclidean continuation;

2) try to fix the parameters by a numerical study of the lattice euclidean
correlation function.

This method would require to deal with finite operator insertions (also for
$\Delta p\ne 0$) to get a smoother extrapolation. This is accomplished
through the same pre-subtraction described in Eq.[\ref{fortyfive}] needed also to eliminate
disconnected contribution, Eq.[{\ref{thirtyfivea}]. Since no chiral extrapolation is needed, this
method could be suitable to the study of B-decays.

\subsection{Parity Conserving Method}

This was the method originally proposed in Ref.\cite{tre}. It is based on the fact that
chiral perturbation theory relates the physical matrix element:
\begin{equation}
\left\langle {\pi ^+\pi ^-} \right|O_{(P.V.)}^{(\pm )}(\mu )\left| {K^0}
\right\rangle =i\gamma ^{(\pm )}{{m_K^2-m_\pi ^2} \over {F_\pi }}\label{fortysix}
\end{equation}
to the matrix element of the parity conserving operator:
\begin{equation}
\left\langle {\pi ^+(p)} \right|O_{(P.C.)}^{(\pm )}(\mu )\left| {K^0(q)}
\right\rangle =-\delta ^{(\pm )}{{m_K^2} \over {F_\pi ^2}}+\gamma ^{(\pm
)}p\cdot q\label{fortyseven}
\end{equation}
$O_{(P.C.)}^{(\pm )}(\mu )$ has a rather complicated ultraviolet structure:
\begin{eqnarray}
& & \tilde {\tilde O}_{(P.C.)}^{(\pm )}(\mu )=\label{fortyeight}\\
& & =Z_{(P.C.)}^{(\pm )}(\mu
a)\left[ {O_{(P.C.)}^{(\pm )}(a)+} \right.\sum\limits_{i=1}^4 {C_i^{(\pm
)}O_{(i)}^{(\pm )}(a)}+\nonumber\\
& & \left. {+(m_c-m_u)C_\sigma ^{(\pm )}O_\sigma (a)+(m_c-m_u){{C_s^{(\pm )}}
\over {a^2}}O_s(a)} \right]\nonumber
\end{eqnarray}
where:
\begin{equation}
O_\sigma =g_0(m_s-m_d)\bar s\sigma _{\mu \nu } G_{\mu \nu }d \label{fortynine}
\end{equation}
and:
\begin{equation}
O_s=\bar s d\label{fifty}
\end{equation}
however, in this case we only have to compute two particle matrix elements (which are easier) and,
therefore, we do not have to worry about final state interaction effects.

\subsection{First Principles}

The distortion of the composite operators in the effective Weak Hamiltonian density
is due to the fact that, in its computation, we insist to integrate up to
zero (lattice) distance.

If we could integrate up to a finite, physical, distance and perform
analytically the final continuum integration the problem could be overcome.

In order to accomplish this, we recall that the Weak Hamiltonian density Eq.[{\ref{one}] is defined through a short distance expansion.

We can therefore directly get the continuum matrix
elements $\left\langle h \right|O^{(i)}(\mu )\left| {h'} \right\rangle$ by studying, numerically  on the
lattice, in the region $a<<\left| x \right|<<\Lambda _{QCD}^{-1}$, the quantity:
\begin{eqnarray}
& & {1 \over 2}\left[ {\left\langle h \right|T[J_L^\mu (x)J_{L\mu
}^+(0)]\left| {h'} \right\rangle +x\leftrightarrow -x} \right]=\nonumber\\
& & =\sum\limits_i {c_i(x;\mu)\left\langle h \right|O^{(i)}(\mu )}\left| {h'}
\right\rangle\label{fiftyone}
\end{eqnarray}
where the coefficientsÊ$\; c_i(x;\mu )$ are related to the $C_i({\mu  \over {M_W}})$ appearing in the
Weak Interaction hamiltonian density by:
\begin{equation}
C_i({\mu  \over {M_W}})M_W^{6-d_i}=\int {dxD(x;M_W)
c_i(x;\mu )}\label{fiftytwo}
\end{equation}
The $C_i({\mu  \over {M_W}})^{'}$s, and therefore the $c_i(x;\mu )^{'}$s, are reliably computable
within perturbation theory. If we use in Eq.[\ref{fiftyone}] the $O(a)$ improved weak currents, then
the whole computation will be automatically $O(a)$ improved.

In the case of $K\to \pi \pi $, the operators contributing in Eq.[\ref{fiftyone}] are $O^{(\pm )}$ and $(m_c^2-m_u^2)O_p$, in the parity violating case, or
$(m_c^2-m_u^2)(m_s+m_d)O_s$, in the parity conserving case.

The perturbative expression of the $c_i(x;\mu )$ is:
\begin{eqnarray}
& & c_i(x;\mu )\propto\label{fiftythree}\\
& & \propto\left( {{{\alpha _s({1 \mathord{\left/ {\vphantom {1 x}}
\right. \kern-\nulldelimiterspace} x})} \over {\alpha _s(\mu )}}}
\right)^{{{\gamma _0^{(i)}} \over {2\beta _0}}}\approx 1+{{\alpha _s(\mu )}
\over {4\pi }}\gamma _0^{(i)}\log (x\mu )+...\nonumber
\end{eqnarray}
where the anomalous dimensions $\gamma _0^{(i)}$ are given by:
\begin{eqnarray}
& & \gamma ^{(+)}=4\quad \gamma ^{(-)}=-8\quad \gamma^{(p,s)} =16
\end{eqnarray}

As with the previous method, no chiral expansion is necessary and therefore it could be applied also to B-decays.

\subsection{Top Quark and CP Violation}

The difficulty here is that the top quark is essential and the GIM mechanism
is not operative.
Even in this case, in analogy with the treatment described in the previous section, it is possible to
work in a world with a light fake top quark in order to fix the matrix
elements of the relevant operators.

In fact we can consider a world in which the top quark has a
fictitious mass $\tilde m_t$ such that ${\raise3pt\hbox{$1$}
\!\mathord{\left/ {\vphantom {1 a}}\right.\kern-\nulldelimiterspace}\!\lower3pt\hbox{$a$}}>>\tilde
m_t>>m_c>>\Lambda _{QCD}$ so that we can choose the subtraction point both in
the region ${\raise3pt\hbox{$1$} \!\mathord{\left/ {\vphantom {1
a}}\right.\kern-\nulldelimiterspace}\!\lower3pt\hbox{$a$}}>>\mu >>\tilde
m_t>>m_c>>\Lambda _{QCD}$ where GIM is operative, and ${\raise3pt\hbox{$1$}
\!\mathord{\left/ {\vphantom {1
a}}\right.\kern-\nulldelimiterspace}\!\lower3pt\hbox{$a$}}>>\tilde m_t>>\mu
>>m_c>>\Lambda _{QCD}$, which mimics the real world.

Identifying the two matrix elements (appropriately evoluted in $\mu $ and
studied as functions of $\tilde m_t$) we can then extrapolate the transition matrix elements to the
real world\cite{sette}.

{\bf Acknowledgements}

My thanks go to C. Dawson, G. Martinelli, G.C. Rossi, C.T. Sachrajda,
S. Sharpe and M. Talevi, with whom most of the consideratons presented
in this talk have been done.

\end{document}